# The effect of novelty on the future impact of scientific grants

Han Zhuang and Daniel E. Acuna[*]
School of Information Studies
Syracuse University

**Abstract**

Government funding agencies and foundations tend to perceive novelty as necessary for scientific impact and hence prefer to fund novel instead of incremental projects. Evidence linking novelty and the eventual impact of a grant is surprisingly scarce, however. Here, we examine this link by analyzing 920,000 publications funded by 170,000 grants from the National Science Foundation (NSF) and the National Institutes of Health (NIH) between 2008 and 2016. We use machine learning to quantify grant novelty at the time of funding and relate that measure to the citation dynamics of these publications. Our results show that grant novelty leads to robust increases in citations while controlling for the principal investigator's grant experience, award amount, year of publication, prestige of the journal, and team size. All else held constant, an article resulting from a fully-novel grant would on average double the citations of a fully-incremental grant. We also find that novel grants produce as many articles as incremental grants while publishing in higher prestige journals. Taken together, our results provide compelling evidence supporting NSF, NIH, and many other funding agencies' emphases on novelty.

## Introduction

According to Thomas Kuhn, science advances more prominently through scientific revolutions than from a slow accumulation of knowledge (Kuhn, 1963). While revolutions in the Kuhnian sense are difficult to define, many researchers believe revolutions come from new results or reorganization of old results in a novel manner (Kindi, 2010). Funding agencies, by and large, recognize that novelty—how unusual a research idea is compared to past ideas—is a necessary component that leads to these revolutions, and these agencies incentivize scientists to pursue novel projects (National Science Board, 2007; NIH, 2016). NSF and NIH, for example, value novelty in their grant submission, instructing applicants to submit "transformative research" (NIH, 2018; NSF, 2019a). The European Research Council (ERC) also favors novelty, stating in their 2016 annual report that investment in "frontier research" enables breakthrough discoveries (ERC, 2017). The National Natural Science Foundation of China (NSFC) follows a similar pattern stating in their manual that "new revolutions in technologies are emerging, and there is a global viewpoint that innovations are driving development..." [translated from Chinese] (NSFC, 2018). But is it possible to relate the novelty of a grant to its future impact? Previous researchers have found a link between publication novelty and future citations (Uzzi, Mukherjee, Stringer, & Jones, 2013; J. Wang, Veugelers, & Stephan, 2017), but grant novelty and citations of publications funded by the grant have not been analyzed to the same extent. There are several difficulties when attempting to make this connection. First, grant citations are not openly available, rendering traditional citation-based novelty quantification techniques impractical. Second, funding agencies only expose a summary of funded grants and not the full text, limiting the granularity of novelty estimation analysis. We hypothesize that grant novelty predicts future citations. Testing this hypothesis is important because grants can give us the first inklings of where impactful research is born, often much earlier than publications can.

---

[*] Corresponding author: `deacuna@syr.edu`

Historically, societies have benefited from scientific novelties that broke away from traditional theories. Isaac Newton's theory of gravity was built from new observations in astronomy (Harper, 2011); Antoine Lavoisier's oxygen theory brought a novel understanding of combustion beyond the phlogiston theory (Le Grand, 1972). Similarly, Albert Einstein's special relativity theory proposed to unify the conservation law of energy and the conservation law of mass into novel coherent principles (Einstein, 2013). The course of history has given novelty its proper credit. At the current pace of scientific knowledge production, however, we cannot afford to wait so long to measure the impact of novel ideas proposed today. With new large, open, and heterogeneous datasets about funding (e.g., https://federalreporter.nih.gov/) and citations (e.g., https://www.crossref.org/), a systematic study of the relationship between funding of novel ideas and their impact seems increasingly feasible.

Several disciplines regularly investigate novelty, innovation, and impact. For example, economics typically studies how patents—a proxy for novelty—impact the economy (Fagerberg & Verspagen, 2009; Martin, 2012). Recently, researchers have examined novelty and impact in science. One of the most prominent techniques to measure novelty is to quantify the unusuality of the citations to previous work. Uzzi et al. (2013) used the atypical combination of reference pairs in publications as an indirect signal of novelty. They found that the most highly cited work has a mixture of traditional and atypical combination of referenced journals. J. Wang et al. (2017) used a similar technique to find that novel publications lead to a higher number of citations in the long-term but to a lower number of citations in the short-term. They also found that novel work tended to be published in lower-tier journals. Recent work by Shi and Evans (2019) also found a relationship between how surprising a piece of work is with the likelihood of ut gaining high number of citations and major awards. Similar results were obtained by Kwon, Liu, Porter, and Youtie (2019) relating technological novelty in publications with future citations. We hypothesize that grant novelty can exhibit a similar relationship to impact and that we can measure this relationship by examining novel combinations of topics in grants' summaries vs the citations to publications funded by those grants.

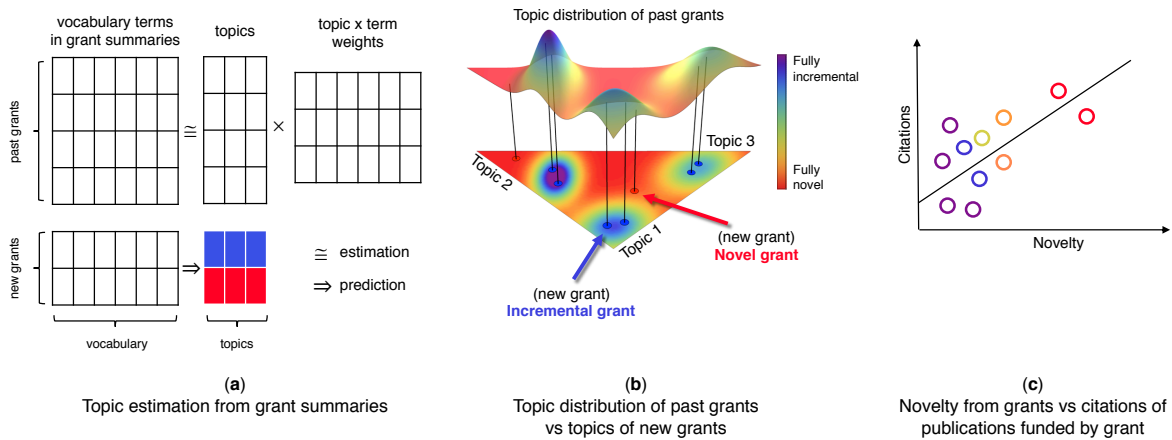

**Figure 1.** A method for estimating the novelty of a grant and its relationship to the citations of the publications funded by the grant. **(a)** For any given year, we estimated the grant topics by using the vocabulary terms in grant summaries of past funded grants. The grant–topic matrix of past grants is decomposed into a grant–topic matrix and a topic–term matrix. The learned transformation from vocabulary to topic is applied to new grants **(b)** Using novelty detection techniques from machine learning, we first estimate the simultaneous topic distribution of past grants. This novelty detection technique predicts the novelty of a new grant. A new grant that is far from the distribution has high novelty and can be labeled as a *novel grant*. Conversely, a new grant that is close to the distribution has low novelty and can be labeled as an *incremental grant*. **(c)** Citations from the publications funded by the grant were used to measure the impact of novelty. If novelty has a positive effect on citations, then a regression analysis that controls for many factors should show a positive marginal effect of novelty on citations.

Novelty is a crucial criterion when evaluating grants yet the link between the novelty and impact of a grant has rarely been explored before. Here, we propose a method that can detect novelty in grant summaries at the time of their submission and relate that to the citation dynamics of publications supported by the grant. Specifically, we attempt to establish this relationship by answering the following two main questions: 1. Do top novel grants produce highly impactful papers? 2. What is the marginal effect of grant novelty on the impact of funded research? We analyze hundreds of thousands of NSF and NIH grants and the funded publications that occurred over nine years. By cross-correlating the measure of novelty with the future citation of publications, novelty shows a significantly positive effect on future impact. This effect remains after controlling for the PIs' experience, the prestige of the journal, year of publication, award amount, and team size. We discuss the implications of this work for funding policy and its limitations.

## Results

In this work, we estimated how novelty of grants relates to their future impact. We used novelty detection techniques from machine learning to measure novelty. Briefly, the technique works by first modeling the topics of grant summaries of the past grants (Fig. 1a, see Methods for more details). Then, a distribution of topics is learned from these past grants (Fig 2b). The likelihood of new grants coming from this distribution determines how incremental (high likelihood) or how novel (low likelihood) they are (Fig. 2b lower simplex). Then, for example, this novelty measure is associated with the citations of the articles funded by the grant (Fig. 2c). If the relationship is positive, novelty is said to be predicted of impact. We also analyzed the effects of novelty on the productivity of grants and the prestige of the journals where funded research is published.

**Novelty Effect on Highly Cited Work:** To investigate the effect of novelty on citation dynamics, we first focused on the most common types of grants for NSF and NIH (see Data section). For NSF, these are all research-focused grants and for NIH, these are the R01, R03, and R21 program grants. Our novelty detection technique (see Methods section) produces a continuous measure of novelty, from 0 (not novel/all incremental) to 1 (all novel/not incremental) by computing how unusual the topics of the grant summary are compared to all grants awarded within the previous two years within the same agency (see Methods section for more details and sensitivity analyses) and therefore can be compared across programs. Importantly, our technique estimates novelty only using the grant summaries—two to three paragraphs describing the problems, objective, and expected results of the grant.

We define highly-novel work as being in the top 10% of novelty across NSF's divisions and NIH's institutes. Similarly, we define highly cited work as citations that are in the top-10% of the citation distribution among funded papers in the same field and year of publication. Previous work has analyzed similar thresholds, including top-1%, top-5% and top-10% (Uzzi et al., 2013; J. Wang et al., 2017) (Our results are essentially similar for top-5% but noisier for the top-1% threshold analysis) Across agencies and years after publication, highly-novel grants have a higher chance of funding research in top-10% of the citation distribution (paired t-test, $t(8) = 6.99$, $p < 0.001$ for NSF and $t(8) = 3.463$, $p = 0.0085$ for NIH, Fig. 2). Overall, these results suggest that highly-novel grants produce significantly more impactful research.

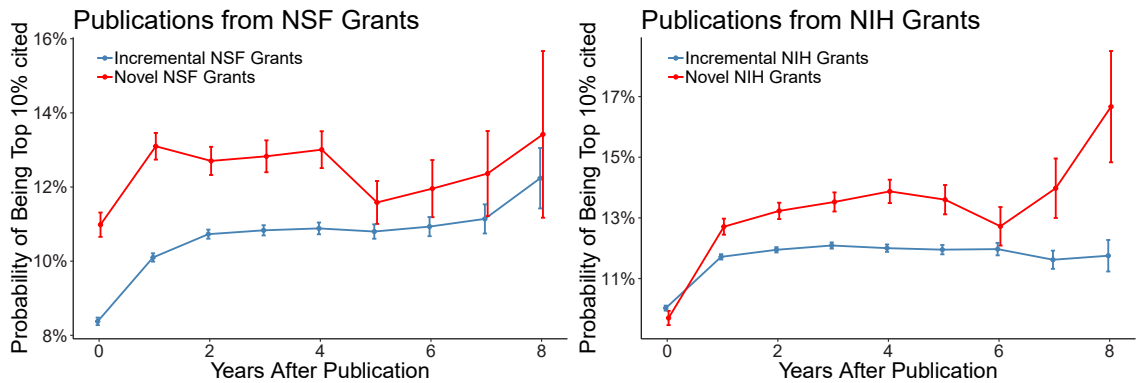

**Figure 2.** Citation dynamics of NSF grants (left) and NIH grants (right). Across years, novel grants (top 10% novelty score), are more likely to produce publications with top 10% citations (within the journal's fields and year of publication). Error bars are standard errors of the mean.

**Novelty Effect on Citation Count:** The above analysis raises a question: how many citation counts are related to a marginal increase in novelty? To answer this question, we use the novelty score as a predictor of the citation count of publications supported by the grant. Considering that many factors can influence the citation count, we use linear regression to control for these other factors. Specifically, we included the prestige of the journal as measured by the Scientific Journal Ranking (SJR) measure (SCImago, 2018), year of publication, the experience of the PI (whether he or she has received grants before), the amount of the award, and team size. The marginal effect of novelty on citation count is significantly positive. The coefficient of novelty score in the regression models quantifies the effect of novelty (Fig. 3).

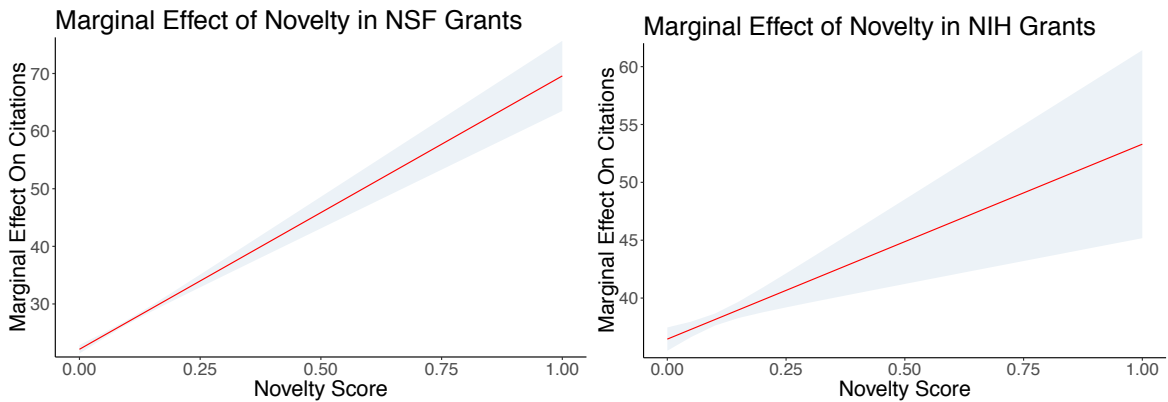

**Figure 3.** The marginal effect of novelty score on citations is significantly positive for NSF (top, $p < 0.01$) and NIH (bottom, $p < 0.01$) with prediction standard error areas. This marginal effect was estimated using a linear regression that controls for the year of publication, PI experience, journal prestige (SJR), award amount, and team size.

For NSF, we found that novelty has a significant positive effect on future citations ($t(127,799) = 13.959$, $p < 0.001$). For NIH, we also found that the novelty score has a significant positive effect on future citation ($t(191,573) = 3.733$, $p < 0.001$). These results suggest that there is a strong positive relationship between novelty and impact, even after considering other potential drivers of this effect. We can perform a rough approximation of the actual citation count effect of novelty on citations. For NSF, a fully novel grant increases citations from 29 to 29+47=76 (a 160% increase) and for NIH, a fully novel grant increases citations from 40 to 40+16=56 (a 40% increase), which across agencies amounts to a doubling of citations.

**Table 1.** Summary of linear regressions for NIH and NSF.

Dependent variable: Citations
Citations ~ Y. of Publication After 2010 + PI Experience + Novelty + SJR + Award Amount + Number of PIs

|  | NSF | NIH |
|---|---|---|
| **Intercept** | 29.797*** | 40.071*** |
|  | (0.866) | (1.785) |
| **Novelty** | 47.453*** | 16.847*** |
|  | (3.423) | (4.579) |
| **Years of Publication After 2010** | -6.439*** | -9.447*** |
|  | (0.0937) | (0.161) |
| **PI Experience** | 0.832 | 0.258 |
|  | (0.594) | (1.203) |
| **SJR** | 7.399*** | 8.043*** |
|  | (0.0625) | (0.061) |
| **Award Amount (in millions of dollars)** | -0.102 | 5.181*** |
|  | (0.152) | (1.020) |
| **Number of PIs** | 1.198*** | 1.936*** |
|  | (0.171) | (0.562) |
| Observations | 127,805 | 191,579 |
| $R^2$ | 0.130 | 0.098 |
| Adjusted $R^2$ | 0.130 | 0.098 |
| Residual Standard Error | 68.333 (df = 127,798) | 115.581 (df = 191,572) |
| F Statistic | 3,182.055*** (df = 6; 127,798) | 3,464.407*** (df = 6; 191,572) |

*$p$<0.1; **$p$<0.05; ***$p$<0.01

Notes: Numbers in parentheses are the standard errors

**Novelty Effect on Prestige of Publication Journal:** Some researchers have shown that there is bias against novelty: highly novel work is published more often in lower-tier journals (J. Wang et al., 2017). To investigate this effect in grants, we compare the ability of top-10% novel grants to produce work in the top-10% SJR journals (SCImago, 2018) within NSF's division and NIH's institute. Our analysis shows that top-10% novel grants in NSF have a significantly higher presence in top-10% SJR journals compared to other grants (+6.1%, paired t-test $t(41) = 4.62$, $p < 0.001$). Similarly, novel NIH grants have significant higher presence in top-10% SJR journals (+5.6%, paired t-test $t(24) = 2.64$, $p = 0.014$). These results suggest that highly novel grants are more likely to produce work published in more prestigious journals.

**Novelty Effect on Productivity of Grants:** Another dimension of scientific impact is productivity. Intuition suggests that novel grants would produce fewer publications because novel ideas are harder to come to fruition. To investigate this hypothesis, we compare the average number of publications of top-10% novel grants with the average number of publications of other grants, within the same grant length and within the same division (in NSF) and Institute or Center (in NIH). The results show that, in NSF, the top-10% novel grants have slightly fewer and non-significant average number of publications than other grants (paired t-test, $t(161) = -0.79858$, $p = 0.4257$). In NIH, top-10% novel grants have a slightly more but not significant average number of publications than other grants (paired t-test, $t(169) = 0.057695$, $p = 0.9541$). This suggests that there seems to be no relationship between the novelty of a grant and the number of publications it produces.

**Novelty and Impact in Different Funding Programs:** The various programs in NSF and NIH have different funding policies. Some of these programs, such as NSF's EArly-concept Grants for Exploratory Research (EAGER), have a high expectation of novelty (NSF, 2009). It would be expected, then, that different programs would also show different effects on impact due to their differences in novelty emphasis. Since the novelty score in this paper is scaled across grants in NSF and NIH, we can compare if one program is more novel than another program, even across agencies (see Methods). To compare the effect of novelty in each program, we computed the probability of their grants being in the top 10% novelty and the probability of their papers being in the top 10% cited among papers in the field within the same year of publication, across programs. This differs from the previous analysis in that here we measured highly cited papers across years. Our results show that the top-3 novel programs are from R01, R03, and NSF normal grants, and top-3 cited programs are R01, NSF's RAPID, and NSF's EAGER (Fig. 4). There is no significant relationship between the average probability of being at 10% in novelty and the average probability of being at the top 10% in citations across programs ($r$= -0.6811534, N = 7, $p$ = 0.09202). For the most common types of grants, NIH's R01 grants have significantly higher probability of top 10% novelty than standard NSF grants ($t(49,354) = 5.5477$, $p < 0.001$) and publications in NIH's R01 grants have significantly higher probability of top 10% citations than publications in standard NSF grants ($t(759,140) = 27.057$, $p < 0.001$). Thus, novelty and impact patterns across programs differ substantially.

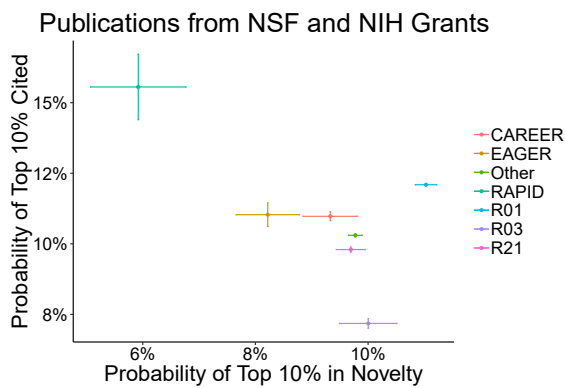

**Figure 4.** Probability of being highly-citations as a function of the probability of being funded by highly-novel grants. The most common NSF funding mechanism has a lower probability of high citations and a lower probability of high novelty compared to the most common NIH mechanism (R01 grant). Error bars are standard errors of the mean.

**Increasing Novelty in NSF and NIH Grants Over Time:** Is the concentration of novel work changing over time? To answer this question, we performed a Pearson's correlation test on the novelty of grants and their year of award, for NSF and NIH separately. Our result shows that from 2010 to 2016 both NSF and NIH show increasing levels of novelty as a function of time (NSF, $r = 0.223$, N = 67,003, $p < 0.001$; $r = 0.157$, N = 43,626, $p < 0.001$). This suggests that there are changes in novelty dynamics over time, and that novelty is increasing.

## Discussion

In this work, we examined whether the novelty of grants is related to the future impact of the articles funded by those grants. We first quantified the relationship between top-novel grants and top-cited publications. A secondary analysis examined the marginal contribution of novelty to citations while controlling for several confounding factors. Finally, we investigated whether novelty relates to the prestige of journals where funded articles are published. In all these measures, we found significant positive relationships, showing that, compared to incremental grants, novel grants produced publications that are cited more often and are published in higher prestige venues. Simultaneously, we found that novel grants do not have different productivity patterns compared to other grants. Taken together, our findings provide robust evidence for preferring the funding of novel research—one of the most prominent science funding policies around the world. Below we discuss implications of our work, its limitations, and transferability to other domains.

*The role of novelty in funding.* Our work supports a positive view of novelty and impact in research, contributing to a rich literature that examines the interplay between science, innovation/novelty, and technology. Science plays an initial and crucial role in technological advances, ultimately driving economic growth (Ch. 6 in Feldman, Link, & Siegel, 2002), and thus forms an important activity for the economy (Stephan, 1996, 2012). Market failures in investments (e.g., see Link & Scott, 2012) paired with wide public support for government involvement (Funk & Rainie, 2015) should incentivize funding of highly innovative and risky scientific projects that would otherwise be discarded by private investments. Recent research has found important links between publication novelty and impact (Kwon et al., 2019; Uzzi et al., 2013; J. Wang et al., 2017). In a recent work by Shi and Evans (2019), the authors developed measures of novelty drawn from surprising (e.g., unpredictable) reuses of journals, conferences, and patents. The authors find that this surprise is related to outsized citations. But evidence for the role of grant novelty in the impact of new knowledge (e.g., publications) has remained elusive. Our results provide robust support for this role while proposing new ways of measuring novelty based on grant text.

*Measuring (grant) novelty based on text.* In our work, we propose a new way of measuring novelty of grants using machine learning on their text. Our novelty estimation technique can have practical value for funding agencies. For example, the method can be applied to conduct systematic novelty estimation for grant applications. Grant reviewers already have an overwhelming task, and automating part of it using novelty measures can accelerate at least some first order ranking and filtering of applications. Also, our new grant novelty measure has some advantages. Previous work on publication novelty has relied on keywords (Azoulay, Graff Zivin, & Manso, 2011) and similarities between those keywords (Azoulay, Stellman, & Zivin, 2006). However, keyword-based metrics do not capture human similarity judgement well because they take time to adapt to changing knowledge landscapes (Achakulvisut, Acuna, Ruangrong, & Kording, 2016), and, more importantly, grants available in FederalExporter do not have keywords at all. Our method uses topic modeling, which captures rich representations of the text (Blei, Ng, & Jordan, 2003), and it is useful for analyzing grant relationships without relying on explicit connections—e.g., there are no direct citations between grants. Our work thus proposes a new approach for measuring novelty in grants by using their text summaries.

*Suggestions for protecting novelty during grant peer review.* There is one aspect that future research should examine: whether the peer review process prevents novel grants from being funded. Previous research has found that peer review has a bias for perceived novelty (Langfeldt, 2006; Linder, Schliwa, Werner, & Gebauer, 2010; Resch, Ernst, & Garrow, 2000). These findings would predict that scientists change their writing patterns to appear more novel than they actually are. However, this would not predict why novel funded grants accrue more citations and fund work published in higher prestige journals— previous research has found bias against novelty in publications (J. Wang et al., 2017). A recent study investigated whether European funding programs supposed to encourage fundamentally risky research do in fact give awards to these kinds of applications more often (Veugelers, Wang, & Stephan, 2019). The authors found that researchers are *penalized* for risk-taking. While this research examined trends in the European context, there is no reason to believe that we would find different results in the U.S. This is somewhat discouraging as the benefits described in our study may get filtered out through the review process. Perhaps other funding mechanisms, such as lotteries (Gross & Bergstrom, 2019), should be considered, lessening this trend.

*Differential novelty across funding programs.* Funding agencies create different funding programs to foster different levels of risk-seeking and success expectations. For example, the Early-concept Grants for Exploratory Research (EAGER) program at NSF aims at funding research especially considered "high risk-high payoff" (NSF, 2019b). However, our results suggest that EAGER grants are not as novel as regular NSF grants but they have a slightly better chance of having more citations than regular grants (Fig. 4). The estimation errors are significant however and our results still hold for regular NSF grants and R01 NIH grants. Interestingly, we found that the Rapid Response Research (RAPID), aimed at funding emerging research with a fast turn-around, is significantly less novel than normal NSF grants but produces significantly more citations. Taken together, these results seem to suggest that it is difficult to motivate grant funding of novel research although it might also indicate that scientists with novel research might still prefer to submit to standard programs because they are better recognized. Previous work has suggested that highly impactful work is randomly distributed across careers (Liu et al., 2018) and that highly accomplished researchers such as Nobel prize winners are similar to other scientists (Li, Yin, Fortunato, & Wang, 2019). Future research could investigate funding-program specific effects on the motivations to modify applicants' behavior and whether there are career-stage effects on novelty.

*Increases in novel research over time.* We find that both NSF and NIH grants became more novel from 2010 to 2016, possibly indicating an increased emphasis on innovation and novelty. One limitation of our investigation is the relatively short time frame of analysis: the citations correspond to data from 2019, and the grant dates are from 2008 to 2016. We only observe the effect of at most 9 years (e.g., accumulated citations through 2019). However, publications might need more time to garner citations, and perhaps a longer time scale could show that incremental grants actually become as cited as novel grants. However, discovering a consistent effect after 9 years of publications would be an encouraging finding for funding agencies—albeit a medium-term effect. Previous work on novelty of publications and impact find that the differences between novel work and not novel are very distinct after 4 years of publication (J. Wang et al., 2017). As the data about grants accumulates, we will have in the future longer time frames to study novelty and impact.

*Generalization to other funding agencies.* There could be several differences between what funders consider transformative research and what we measure here. For example, the Department of Energy's Advanced Research Projects Agency Energy (ARPA-E) prioritizes community creation around certain topics, suggesting that citations and bibliometric impact operate only secondarily. Many DARPA projects are started by defining "end-game" scenarios—ideals that are almost impossible to achieve—that trigger innovative problem solving (Dubois, 2003). Thus, citation and impact may play little role in DARPA's funding decision. Similarly, some private foundation might not be in a position to take on risky research and therefore like to fund safer incremental projects. Thus, our results are not necessarily relevant for all funding scenarios.

*Effects in interdisciplinary research.* Interdisciplinary research seeks to combine ideas from several fields and therefore would naturally sit at the intersection of disciplines (Engineering Committee on Science et al., 2004), and could appear more novel. Previous work has found that atypical combinations are highly correlated with interdisciplinarity (Boyack & Klavans, 2014). In the work proposed by Staudt et al. (2018), the authors proposed several measures of novelty and impact based on text and citation data, and found a general relationship between the two in scientific publications. In our results, however, we are controlling for some kinds of interdisciplinary effects by using the SJR measure of impact of a journal and the team size (Table 1).

*Data quality issues.* We of course can only measure the effects on publications that were properly reported to NSF and NIH. There could be publications that were funded by these grants but the authors chose not to report them or there were manuscripts published too late, after the end of the grant and hence not reported. In the FederalExporter datasets, there are around 14 US public funding agencies, and they may vary in how they track funded publications. We found that NSF seem to have issues tracking past publications compared to NIH. In the dataset section, we describe how we found a smaller proportion of the publications associated with NSF compared to NIH. In the future, we will investigate other funding acknowledgement services such as those provided by Dimensions (https://dimensions.ai), the Web of Science API, or Scopus API. Many of these services, however, already largely rely on FederalExporter.

*Limitations of using only funded research.* The grants that we are analyzing have already been filtered by the peer-review process. Perhaps peer reviewers are selecting novel grants that are likely to be impactful. Previous evidence, however, has shown that grant reviewers do not agree well on their decisions (Pier et al., 2018), and reviewers are not good at predicting future impact (some evidence from Sociology in Teplitskiy & Bakanic, 2016). However, the evidence is not so conclusive on whether reviewers predict citations. A study by Barnett, Glisson, and Gallo (2018) found no predictability whereas Fang, Bowen, and Casadevall (2016) and (Kaltman et al., 2014; Lauer, Danthi, Kaltman, & Wu, 2015) found that better review scores led to higher citations in NIH. To examine this and other possibilities, we hope to analyze unfunded grants and whether unfunded researchers who continue with the research anyway get more citations in spite of rejection. Recent work by Y. Wang, Jones, and Wang (2019) showed evidence that early rejections in the grant review process has long-term positive effects on citations. Alternatively, we could apply more sophisticated, matching methods to get at the causality of novelty over citation—e.g., Imbens and Rubin (2015). In our results, we have taken precautions to control for factors that we think could be confounds of the analysis, and we suspect that the same results would be present before grant peer review.

*Challenges when measuring novelty in grants.* We are measuring novelty using unsupervised machine learning, which is hard to validate (James, Witten, Hastie, & Tibshirani, 2013). Most previous research on novelty and innovation detection also rely on unsupervised learning techniques (James et al., 2013; Rzhetsky, Foster, Foster, & Evans, 2015; Veugelers et al., 2019; J. Wang et al., 2017). Our results, however, hold under several parameter modification and outlier thresholds ("pollution fraction" parameter in Methods section). This suggests that the method is being at least highly consistent in what it considers novel. Future research could consider supervised datasets where experts determine what were truly novel grant proposals. For example, access to full NIH applications could allow us to analyze the innovation section of the application and the innovation assessment by the review panels. Thus, our novelty machine learning detection method is robust but could be improved with ground truth data.

*Alternative measures of impact.* We measure impact by looking at the citations of the articles funded by the grant. But it is possible that grants that are more incremental produce effects on other publications that are not acknowledged as being funded by the grants, and whose combined number of citations is higher than the publication acknowledged by the grant. Also, the impact of the grant does not necessarily only relate to citations. Recently, other forms of impact have been measured such as patents, software, datasets (Zeng, Shema, & Acuna, 2019), and social network discussions (e.g., "alt-metrics" effects in Thelwall, Haustein, Larivière, & Sugimoto, 2013). Citations are still one of the primary ways of measuring impact in scientific research, and thus our results provide relevant evidence relating novelty to impact, albeit limited.

## Conclusion

This study found important evidence for the link between grant novelty and the impact of publications funded by grants. The most important funding agencies in the US, Europe, and China explicitly state the benefits of innovative research yet evidence between innovation and funded research's impact has been elusive. Our results showed that grants with novel research fund articles with significantly higher citations, published in significantly more prestigious journals. We also found novelty and citation differences across funding programs. We related our results to grant peer review, tracking of funded research, interdisciplinary research, and discuss limitations of our findings.

We believe that funding policy research has an almost unprecedented opportunity to understand how knowledge production is related to impact. This is in part thanks to new datasets about funding and publications, and the marriage between computational power and increasingly accurate and interpretable machine learning techniques to measure countless aspects of funding. We hope that our systematic analysis of novelty enables new complimentary information used during grant peer review, and that new incentives are created to promote the submission and acceptance of innovative grants. We think that this would unlock significant improvements to how science plays a role in technological advances and general improvements to the welfare of society.

## Methods

**Dataset.** We collected NSF and NIH grants from 2008 to 2016 from Federal Exporter (NIH, 2019). There are 111,764 NSF grants within this period. There are some duplicate grants, and only the earliest versions of them were kept in the analysis. This duplication removal left us with 92,355 NSF grants. Additionally, there are some non-research grants in the data, such as I-Corps grants and workshops. We used active learning techniques to create a logistic regression classifier for non-research grant detection. More specifically, we first created a list of non-research grants from I-Corps and workshop grants and then a list of random grants. We labeled the above two lists to create an initial training data set. We then performed 10 rounds of iterations of active learning to label new grants based on their language usage. The average Area Under the Curve (AUC) of the classifier in three-fold cross-validation was 0.93 (± 0.03). We removed all the non-research grants from the data with this logistic regression model, which left us with 91,162 NSF grants.

There are 672,286 NIH grant records in Federal Exporter within the 2008 to 2016 period. We studied R01, R21 and R03 grants because they are the most common grants in NIH. There are 297,495 grant records from the above three programs. We removed continuously funded NIH (repeated) grant records by only keeping a copy of the earliest version. This preprocess left us with 86,862 unique NIH grants.

NIH and NSF have the publication records of their grants available to the public. We used Scopus API to collect the total citation count for these publications and other metadata. We found 270,958 articles of the 493,808 articles associated with grants from NSF and 651,335 of the 751,062 articles associated with grants from NIH. In addition, we used Open Citation (OpenCitations, 2019) data set for the citation dynamics analysis. We found all Scopus-matched NSF papers (270,958) in Open Citation and 506,303 papers in Open Citation data set of the 651,335 Scopus-matched NIH publications.

To investigate the effect of novelty on impact, each grant was associated with a novelty score (a proxy for novelty) and the citations of the grant's resulting publications. The novelty score is quantified as a number between 0 and 1 (0 refers to the least novel, and one refers to the most novel) and is, based on the grant summaries of grants.

**Quantification of novelty:** The novelty score is used as a proxy of how novel a grant is. Since novelty tends to be different from existing knowledge or a combination of old knowledge in a novel manner, novelty can be estimated by comparing a grant to grants awarded in the past. These new ideas and new approaches can be quantified via grant summaries. We now describe how such a procedure is performed.

We first estimate the term frequency-inverse document frequency (tf-idf, Eq. 1) (in unigrams and bigrams) for grant summaries in a window of past years (two years in our main results). Since tf-idf considers the document frequency, each term in the tf-idf vector has a weight (term frequency in this document divided by how often this term appears across the documents). The weights in tf-idf vectors are informative because they tell us how unusual the word is with respect to the rest of the documents.

$$\text{tf-idf}(term, document) = \text{tf}(term, document) \times \text{idf}(term) \qquad (1)$$

Since the tf-idf vectors have high dimensionality (e.g., number of terms in all the grants), we reduce this dimensionality using a technique known as non-negative matrix factorization (Cichocki & Phan, 2009; Fevotte & Idier, 2011). This is, we represent the matrix of tf-idf of all grants (*V* is a number of documents by number of terms matrix in Eq. 2) as a multiplication of a tall and skinny matrix of *topic* loads (*W* in Eq. 2, with 50 dimensions in our main results) and a short and wide matrix of topic to term weights (*H* in Eq. 2, discarded in our study).

$$V = WH \tag{2}$$

After the language of grant summaries is represented with *W*, we use a one-class Support Vector Machine (One-Class SVM) (Schölkopf, Platt, Shawe-Taylor, Smola, & Williamson, 2001) to learn the distribution of topics of past grants and estimate the novelty of current grants. Concretely, the method finds a set of *support vector* grants, which are representative of past grants, by solving the following mathematical programming formulation

$$\min_{w \in F, \xi \in R^l, \rho \in R} \quad \frac{1}{2}\|w\|^2 + \frac{1}{vl}\sum_i \xi_i - \rho \tag{3}$$

$$\text{subject to } (w \cdot \Phi(x_i)) \geq \rho - \xi_i, \xi_i \geq 0 \tag{4}$$

$$f(x) = sgn\left((w \cdot \Phi(x)) - \rho\right) \tag{5}$$

where $w$ is the normal vector to the hyperplane, $\nu$ is the upper bound of the fraction of outliers, $l$ is the number of observations, $\xi_i$ is a nonzero slack variable, $\rho$ determines the offset from the hyperplane to origin, and $\Phi$ is the kernel function. Equation (3) and (4) are the loss function of one-class SVM, and Equation (5) determines if a data point is an outlier, by computing the data point's distance to the hyperplane.

The learned One-Class SVM is then used to estimate novelty as follows. For a new grant, $x'$, the quantity $(w \cdot \Phi(x')) - \rho$ in Eq. 5 is computed. This quantity represents the raw distance of the grant to the boundary of the distribution of past grants and therefore gives us a natural measure of continuous novelty. This measure, however, is not directly interpretable because it is in terms of an underlying monotonically-increasing function (like a cumulative distribution function). Therefore, we rescale novelty scores using a min-max scaling across all agencies and years. The final novelty score is 1 for the most novel grants, and 0 for the most incremental grant.

**Sensitivity analysis of novelty detection:** We perform two control analyses: First, we verified that novelty diminishes if grants from today were "cloned" into the past, therefore decreasing the likelihood that the ideas of today are truly novel. Second, we verified that the results were similar if the window of analysis varied and the parameters of the topic model and novelty detection technique were modified as well.

To analyze the robustness of novelty detection with real data, we perform the following control. We used one grant to estimate how much novelty decreases if past grants were to be increasingly similar to today's grants. We used grant number "0969857", "LIGO Observations and Gravitational Wave Astronomy" (NSF) as the probe grant. We then added a noisy version of the grant using its term-frequency vocabulary plus Gaussian noise. As we increased the number of these noisy versions we added to the past, we saw a decrease in the novelty measure. The novelty measure dropped from 0.516 to 0, after we replaced 4% of previous grants with the same number of similar grants to grant number "0969857". These tests show that novelty detection can detect past grants that are similar.

We then analyzed the robustness of the topic model and novelty detection parameters. We examined different "pollution fractions", the $\nu$ parameter ($\nu = 0.01$ and $\nu = 0.1$). This parameter is the upper bound on the fraction of outliers and controls the minimum complexity of the distribution (Schölkopf et al., 2001). The novelty (from the analysis with a $\nu$ of 0.1 and 0.01) still has a positive effect on the citation count in our linear regression (first two rows of Table 2). We also examined the effect of the number of topics on the novelty detection method, and the linear regression results. We found that changing the number of topics from 50 to 100 did not change the direction of the results (third row of Table 2). Finally, we examined the effect of different window sizes (WS) from using one previous year to three previous years. Again, the results of the linear regression did not change direction (fourth and fifth rows of Table 2).

Table 2. Summary of sensitivity analysis. Across a range of parameters for the topic model and novelty detection, the novelty coefficient remains positive and mostly significant for NSF and NIH ($\nu$= the upper bound of outlier fraction and controls the minimum complexity of the distribution; TN: number of topics; WS: widow size in years)

|  | NSF | | NIH | |
|---|---|---|---|---|
|  | Novelty coefficient | p-value | Novelty coefficient | p-value |
| $\nu = 0.01$<br>TN = 50<br>WS = 2 | 39.00 | <0.001 | 14.80 | <0.001 |
| $\nu = 0.1$<br>TN = 50<br>WS = 2 | 50.25 | <0.001 | 16.21 | <0.001 |
| $\nu = 0.05$<br>TN = 100<br>WS = 2 | 75.04 | <0.001 | 6.43 | 0.20 |
| $\nu = 0.05$<br>TN = 50<br>WS = 3 | 79.72 | <0.001 | 12.75 | 0.103 |
| $\nu = 0.05$<br>TN = 50<br>WS = 1 | 24.79 | <0.001 | 18.90 | <0.001 |

## Acknowledgments

H. Zhuang and D. E. Acuna were partially funded by NSF Awards #1646763 and #1800956. We thank David C. Popp for comments on the draft. We thank James A. Evans for sharing a preliminary version of his manuscript "Science and Technology Advance through Surprise".